\documentclass[12pt,preprint]{aastex}

\shortauthors{McCabe, Duch\^{e}ne \& Ghez}
\shorttitle{NICMOS Images of GG Tau}

\begin{document}

\title{NICMOS images of the GG Tau Circumbinary Disk}
\author{C. McCabe, G. Duch\^{e}ne \& A.M. Ghez}
\affil{Department of Physics and Astronomy, University of California, 
Los Angeles}
\authoraddr{405 Hilgard Ave, Los Angeles, CA 90095-1562}
\email{mccabe@astro.ucla.edu}

\begin{abstract}
We present deep, near-infrared images of the circumbinary disk
surrounding the pre--main-sequence binary star, \objectname[]{GG Tau}
A, obtained with NICMOS aboard the Hubble Space Telescope.  The
spatially resolved proto-planetary disk scatters $\sim$1.5\% of the
stellar flux, with a near-to-far side flux ratio of $\sim$1.4,
independent of wavelength, and colors that are comparable to the
central source ($\Delta\left(M_{F110W}-M_{F160W}\right) = 0.10 \pm
0.03 $, $\Delta\left(M_{F160W}-M_{F205W}\right) = -0.04 \pm 0.06 $);
all of these properties are significantly different from the earlier
ground-based observations.  New Monte Carlo scattering simulations of
the disk emphasize that the general properties of the disk, such as
disk flux, near side to far side flux ratio and integrated colors, can
be approximately reproduced using ISM-like dust grains, without the
presence of either circumstellar disks or large dust grains, as had
previously been suggested.  A single parameter phase function is
fitted to the observed azimuthal variation in disk flux, providing a
lower limit on the median grain size of $a > 0.23$ \micron. Our
analysis, in comparison to previous simulations, shows that the major
limitation to the study of grain growth in T Tauri disk systems
through scattered light lies in the uncertain ISM dust grain
properties. Without explicit determination of the scattering
properties it is not possible to differentiate between geometric,
scattering and evolutionary effects. Finally, we use the 9 year
baseline of astrometric measurements of the binary to solve the
complete orbit, assuming that the binary is coplanar with the
circumbinary ring. We find that the estimated 1$\sigma$ range on disk
inner edge to semi-major axis ratio, $3.2 < R_{in}/a < 6.7$, is larger
than that estimated by previous SPH simulations of binary-disk
interactions.
\end{abstract}


\section{Introduction}
GG Tau A is one of the few T Tauri systems in which a disk has been spatially
resolved at both millimeter and near-infrared wavelengths and
is therefore an ideal system for detailed studies of proto-planetary disk geometry and
composition.
As the second brightest millimeter object in the survey of Beckwith et
al. (1990), GG Tau was an early target for millimeter wave
interferometry studies (Simon \& Guilloteau 1992; Kawabe et al. 1993;
Dutrey, Guilloteau \& Simon 1994; Guilloteau, Dutrey \& Simon 1999,
hereafter G99). These observations showed the emission to be from a
massive ($0.13 M_{\odot}$) disk surrounding GG Tau A, the closest pair
of stars in the GG Tau quadruple stellar system\footnote{GG Tau A-B,
Aa-b and Ba-b are separated by 10\arcsec, 0\farcs25 and 1\farcs48
respectively (Leinert et al. 1991).}, with the bulk of the emission
arising from a distinct ring structure extending from 180 to 260 AU
(assuming a distance of 139 pc; Bertout, Robichon \& Arenou,
1999). The circumbinary disk was subsequently detected via
near-infrared scattered light in multi-wavelength, ground-based
adaptive optics (AO) images (Roddier et al. 1996, hereafter known as
R96) and in space-based Hubble Space Telescope (HST)
1 \micron\ polarimetric images (Silber et al. 2000) and WFPC2 optical images
(Krist, Stapelfeldt \& Watson 2002). 

Analysis of the ground-based near-infrared images resulted in disk
colors that appear to be redder than the central stars (R96). Such a
red color excess could either indicate that the circumbinary disk has
a substantial population of large dust grains, or that circumstellar
disks, which are coplanar with the circumbinary disk, redden the
stellar light before it scatters off the circumbinary disk.  Comparing
the variation of disk magnitude with color around the disk, Roddier et
al. find a trend for fainter regions of the disk to be redder, leading
them to suggest that circumstellar disks are the cause of the observed
red color excess.  The existence of circumstellar disks around the
individual components in GG Tau A is supported by observations of
near-infrared excesses and strong hydrogen emission lines in each
component (White et al. 1999). No information, however, is currently
available regarding their orientation and in fact many disks in binary
systems show evidence of non-coplanarity (e.g., Stapelfeldt et
al. 1998; Monin, M\'enard \& Duch\^ene, 1998; Jensen et al. 2000; Wood
et al. 2001). To test the circumstellar disk hypothesis, Wood, Crosas
\& Ghez (1999) ran Monte Carlo scattering simulations of the GG Tau
circumbinary ring with and without circumstellar disks being
present. Using the Kim, Martin \& Hendry (1994) interstellar medium
(ISM) grain properties and the disk geometry derived by Dutrey et
al. (1994), they find that the circumstellar disks are required to
match the disk to star flux ratios and near side to far side flux
ratios observed in the ground-based AO images.  While these
simulations roughly reproduce the estimated quantities due to the
large observational uncertainties, they predict blue rather than red
disk color excesses. In order to make further progress in
understanding the role of circumstellar disks and the possibility of
large particles, much higher signal to noise measurements of the
circumbinary disk are required at multiple wavelengths.

In this paper we present new high angular resolution, multi-wavelength
observations of the GG Tau system taken with NICMOS aboard HST. These
deep, space-based observations offer a more stable point spread
function (PSF) than that obtainable with ground-based AO systems,
allowing a more accurate determination of the disk properties.  The
observations are outlined in \S\ref{obs} and the data reduction and
PSF subtraction method used are explained in \S\ref{reduc}. Newly
derived stellar and circumbinary disk properties are presented in
\S\ref{res} and are discussed in sections \S\ref{discus} and
\S\ref{implications} in light of additional Monte Carlo simulations.

\section{Observations}
\label{obs}
GG Tau A ($\alpha = 04^{h} 32^{m} 30^{s}.3$, $\delta =
+17^{\circ} 31^{'} 41^{''}, J2000$) was observed using the near-infrared
camera, NICMOS, aboard HST on 1997 October 10.  Deep images were
obtained with the F110W ($\lambda_{o}=1.03 \micron$,
$\Delta\lambda=0.55 \micron$), F160W ($\lambda_{o}=1.55 \micron$,
$\Delta\lambda=0.40 \micron$), and F205W ($\lambda_{o}=1.90 \micron$,
$\Delta\lambda=0.60 \micron$) filters\footnote{The NICMOS F110W and F160W 
filters correspond well to the standard ground-based 
J and H band filters, with the F160W
filter providing the closest match. The F205W filter is centered $\sim$0.3 \micron\ 
shorter than the standard K band filter and is $\sim$0.2 \micron\ wider.}. 
In the F110W and F160W filters,
GG Tau was imaged 9 times on NIC1, the NICMOS camera with the smallest
field of view, 11$\arcsec \times$ 11$\arcsec$, and pixel scale of
0.$\arcsec$043 per pixel (Thompson 1995).  Each image was offset by
2.$\arcsec$2 in a spiral dither pattern in order to reduce residual
flat field uncertainties and the dither size was set to a non-integral
number in order to reduce intra-pixel sensitivity variations, while
also keeping the circumbinary disk in the field of view.  Each
individual dither image was integrated for 128 seconds in the
multiaccum mode.  The F205W filter data were taken using NIC2, which
has a 19.$\arcsec$2 $\times$ 19.$\arcsec$2 field of view and a pixel
scale of 0.$\arcsec$075 per pixel. Due to the significant thermal
background at this wavelength a dither-chop observing pattern was
employed, whereby each image is composed of one 128 second integration
on GG Tau followed by a 32$\arcsec$ chop and a 128 second integration
on a blank piece of sky. Nine images were obtained in this manner.

Because the aim of our research was to detect faint, diffuse emission
close to bright sources, a good characterization of the point spread
function was required. To this end, images of a calibrator star,
NTTS 042417+1744, were obtained in the same manner as GG Tau in each
of the above mentioned filters. This weak-lined T Tauri star has a number
of properties which were used to select it as a calibrator: no
infrared excess indicative of circumstellar material (Walter et
al. 1988), known to be single from previous speckle interferometry
observations (Ghez et al. 1993), only half a magnitude brighter than
GG Tau Aa in the J band (1.25 \micron), and a spectral type of K1,
which is similar to the stellar components of the binary star (K7 and
M0.5 for GG Tau Aa and Ab respectively; White et al. 1999).

\section{Data Analysis}
\label{reduc}
\subsection{Image Processing}
Initial data reduction was carried out by STScI, providing a quick
analysis of the data quality. The images were recalibrated with the
latest calibration reference files using the STScI's pipeline routine
CALNICA (Bushouse, Skinner \& MacKenty, 1996), which performs standard
image reduction tasks: removal of bias and dark current levels,
interpolation over known bad pixels, flatfielding, and removal of all
cosmic ray hits. Once the data were calibrated, each individual image
was sub-pixelated by a factor of 2.

In order to obtain estimates of the properties of the central binary
(\S\ref{orbit}) and the circumbinary disk (\S\ref{cbdisk}), a model of the central
binary star was generated through PSF fitting and then subtracted from
the individual images. Both the empirical PSF (observations of NTTS
042417+1744) and TinyTim model PSFs (Krist, 1993) were used. The
PSF-fitting was carried out for each of the nine GG Tau images by
minimizing the chi-squared between the PSF model and the data in the
regions dominated by the binary star (the central area of $\sim 0.67\
arcsec^{2}$) and, in the case of the empirical PSF, the diffraction
spikes (an additional $\sim 1.65\ arcsec^{2}$ comprising 4 rectangular
regions covering the diffraction spikes beyond 2\arcsec distance from 
the central binary).  For the empirical PSF, all nine PSF estimates 
were evaluated and the one that produced
the best fit, i.e., the smallest residuals, was selected for each GG
Tau image. Although the TinyTim routine allowed us to match the colors
of the two GG Tau A stellar components as well as the focus position
of the observations (see \S3.2), the empirical PSF models consistently
produced smaller residuals than those created using TinyTim, and thus
the final values for the central binary and the circumbinary disk were
derived using the empirical PSF model subtracted images.  The
difference in the quality of the fits between the empirical PSF and
the TinyTim PSF can be understood in terms of (a) small scale PSF
variations due to the telescopes ``breathing'', the change in the
focus of the telescope on timescales of an orbit due to a varying
thermal load, and (b) the presence of scattered light off the
telescope optics (Krist and Hook, 1999).  Neither of these effects are
incorporated into the TinyTim PSF models used, and both are measured
with the empirical PSF.  The nine model subtracted images were
combined together to form an average subtracted image at each
wavelength, which are shown in Figure~\ref{mosaic}. Azimuthal averages
of the model subtracted images are plotted in
the first row of Figure~\ref{profile}.

\subsection{Bias and Uncertainty Estimates}

\subsubsection{Measurement Uncertainties}
A map of the measurement uncertainties was created by taking the
standard deviation of the mean from all nine PSF subtracted images of
GG Tau. Azimuthal averages of these statistical uncertainties are
shown in the second row of Figure 2; while these
values provide a good estimate of the uncertainties associated with
the measurement and subtraction process, they do not take into account
the presence of any systematic effects.

\subsubsection{Systematic Biases and Associated Uncertainties}
\label{bias}
Two systematic effects that alter the structure of the PSF have been
identified.  The first, which affects all three band passes, arises
because the observed PSF, NTTS 042417+1744, is slightly bluer than
either component of GG Tau. The second is an inadvertent focus offset
between the object and PSF in the F205W data.  These effects create a
systematic bias and an additional source of uncertainty, both of which
were modeled and accounted for.
 
In order to investigate the consequences of the small color
differences between the object and PSF, we generated polychromatic
models of GG Tau Aa, Ab and NTTS 042417+1744 using TinyTim, matching
the spectral shape of these objects using previous photometry from
Ghez et al. (1997) and Walter et al. (1988).  The PSF fitting and
subtraction process was run on the TinyTim model images in order to
produce a model of the bias term (i.e., the subtraction residuals
caused by this color mismatch); azimuthal averages of this color bias
are plotted in the first row of Figure~\ref{profile}.  To
obtain an estimate of the uncertainty associated with this measurement
of the color bias, the method was repeated using a PSF that varied in
spectral type by the assumed uncertainty in the calibrators spectral type
(one subclass).  The difference in residuals between using a PSF of
spectral type K0 and K1 is our estimate of the statistical uncertainty
on the color bias term (plotted in the second row of
Figure~\ref{profile}); the uncertainty on the color bias is
significantly smaller than the measurement uncertainties.

While this method was sufficient for the F110W and F160W data, the
F205W data required an additional systematic effect to be accounted
for.  Observations of GG Tau on NIC2 were run in parallel mode with
polarization observations (which are not presented here) on NIC1.  Due
to the unexpected thermal problems, each NICMOS camera had a different
optimal focus position and therefore parallel observations were taken
at non-optimal focus positions.  Since these observations were
scheduled prior to the recognition of this problem, the polarization
data were arbitrarily regarded as the prime observation. Therefore
both the polarization and F205W observations were made with the NIC1
focus setting, resulting in slightly out of focus F205W
observations. Unfortunately, the F205W data on the PSF were taken in a
separate orbit with no parallel observations and therefore at the
optimal NIC2 focus setting. This mismatch in focus between the object
and calibrator in the F205W data set was quantified using the same
TinyTim modeling used to determine the color bias in the F110W and
F160W data, but incorporating both the color difference and mean focus
offset. As shown in the third column of Figure~\ref{profile}, the
focus bias is the source of a considerably large systematic error.  To
estimate the systematic uncertainty associated with the measurement of
this bias, a polychromatic TinyTim model of GG Tau was created with
the focus set at the {\it maximum} focus position of the data set, and
then subtracted using a K0 type PSF set at the {\it mean} focus
position of the PSF data set. As focus and color mismatches can cause
similar problems, this model set-up maximizes the difference between
the PSF for the object and calibrator data sets. In contrast with the
F110W and F160W data sets, the uncertainty associated with the bias
term in the F205W data dominates the measurement uncertainties.

\subsubsection{Combining Statistical and Systematic Effects}

All of the results presented in \S\ref{res} are obtained from analysis of
images that have had estimates of the systematic bias subtracted out
and uncertainties that have been combined in quadrature, as described
in equations \ref{eq:final} and \ref{eq:errors}.

\begin{equation}
Final_{\lambda} = Img_{\lambda} - Bias_{\lambda}  \label{eq:final}
\end{equation}

\begin{equation}
\sigma^{TOT}_{\lambda} = \sqrt{\left( \sigma^{img}_{\lambda}\right)^{2} + 
\left( \sigma^{bias}_{\lambda}\right)^{2} }    \label{eq:errors}
\end{equation}

\noindent $Img_{\lambda}$ is the median of the nine PSF subtracted
images and $Bias_{\lambda}$ is the systematic bias from the mismatch
in color and focus position. The two error terms in
equation~\ref{eq:errors} are the statistical uncertainties of the
image measurement and bias modeling respectively. The final row in
Figure~\ref{profile} displays the azimuthally average of the bias
corrected signal-to-noise (e.g., $Final_{\lambda} /
\sigma^{TOT}_{\lambda}$). Analysis of the F205W dataset is
limited to the measurement of the binary star
properties (\S\ref{orbit}) and integrated disk photometry (\S\ref{colors}), since
it is significantly affected by the focus bias. Although this bias
and its associated uncertainty
have been modeled, the models were
made using TinyTim PSFs that have been shown to deviate somewhat 
from the observed PSFs.

\section{Results}
\label{res}
In this section we present the observed properties of the stellar
system and the circumbinary disk. The binary orbital motion has been
followed for the past 9 years and the newly estimated disk mass and
inclination (G99) allow us to derive estimates of all the orbital parameters for
the binary system using only minimal assumptions (\S\ref{orbit}).
For the circumbinary disk (\S\ref{cbdisk}), we focus on the
following observational quantities that have, in the past, been considered 
useful tools in comparing observations and disk
models: in \S\ref{1-D}, the disk morphology (Wood et al. 1999), in
\S\ref{colors}, the disk to star flux ratio (Wood et al. 1999) and
color excesses (R96), and in \S\ref{azvar}, the spatially resolved
color-magnitude relationships (R96) and near side to far side flux
ratios (Close et al. 1998; Wood et al. 1999).  These properties are
then compared to earlier measurements. Deviations from previous
estimates of these quantities are most likely attributable to a
combination of low SNR and over-deconvolution of the disk in the more
challenging AO observations.

\subsection{Central Binary Star Properties}
\label{orbit}

Table~\ref{astrom} lists the values for the binary separation,
position angle (P.A.) and magnitude of each component obtained from
the PSF fitting of the central binary star. The uncertainties are the
standard deviation of the estimates obtained from the nine individual
measurements, plus additional contributions from the uncertainties of
the spacecraft orientation (0\fdg03; Holtzmann et al., 1995a) and the
pixel scale (0.5\%; NICMOS Instrument Handbook). The 5\% uncertainty
in the absolute photometry calibration, however, is not included as
our analysis rests on the relative photometry of the disk compared to
the central stars. The reported individual stellar flux density
uncertainties are dominated by PSF variations. Wide aperture
photometry was carried out with the assumption that the disk is a
small contamination on the stellar light ($\sim$ 1\%, see
\S\ref{colors}); this both checks the PSF fitting measurements and
provides a more precise value of the total stellar flux which is used
in \S\ref{colors}. Using an aperture 2\farcs6 in radius, we obtain a
total binary flux density that is consistent to within 1$\sigma$ with
the values found through PSF fitting (see Table~\ref{astrom}).

Over the 9 year period 1990 to 1998, the binary system was repeatedly
observed (Leinert et al. 1993; Ghez et al. 1995, 1997; Roddier et
al. 1996; Silber et al. 2000; White \& Ghez 2001; Woitas, K\"ohler \& Leinert 2001; Krist et
al. 2002). Both the new HST photometric and astrometric measurements
agree well with the other reported measurements. Furthermore, the
astrometric collection of measurements has grown significantly such
that estimates of the relative velocities in the plane of the sky can
be greatly improved. Combining the HST measurements with those
published previously\footnote{Average uncertainties have been assigned
to all measurements obtained with the same method. As a result,
decreasing weights are assigned to HST ($\sigma_{r} = 2.5 mas$), speckle
interferometry ($\sigma_{r} = 5 mas$) and adaptive optics ($\sigma_{r} =
10 mas$) astrometric results, respectively. Furthermore, the astrometry
from Leinert et al. (1993), obtained from three separate 1D
measurements is not included in our fit.} and assuming constant linear
motion, we find a velocity of $6.68 \pm 0.52$ mas yr$^{-1}$ at a position
angle of $260^{\circ} \pm 5^{\circ}$, with an average relative
position of $0\farcs2485 \pm 0\farcs0013$ at the position angle of
$358\fdg2 \pm 0\fdg3$. This velocity has a factor of 5 smaller
uncertainties than previous estimates (Ghez et al. 1995; Woitas et al. 
2001), mostly resulting from including more
measurements. With the greater precision in both the velocity reported
here and the total mass of the system, derived from the rotation of
the circumbinary disk (G99), it is now possible to solve for the whole
orbit, assuming only that the orbit is coplanar with the circumbinary
ring, and that the ring is intrinsically
circular. Appendix~\ref{getorb} provides the details of how the
orbital parameters, $e$, $a$, $P$, $T_{0}$, $\omega$ and their
uncertainties are derived from the observables,
$\overrightarrow{r}_{2D}$, $\overrightarrow{v}_{2D}$, $M$, $i$, $D$,
$\Omega$.

We find that the stars are in an elliptic orbit with an eccentricity
$e=0.32\pm0.20$ and a semi-major axis $a=35^{+22}_{-8}$\,AU. The
corresponding orbital period is $185^{+195}_{-55}$\,yr. We emphasize
that the uncertainties quoted here include the uncertainty on the
distance to the Taurus molecular cloud, which must also be taken into account when considering the
system mass (G99). Indeed, together with the uncertainty in the
measured velocity, this is the dominant source of error in the quoted
uncertainties. Although the orbital parameters derived here still have
large uncertainties, this analysis shows the applicability of this
method. Additional astrometric data over the next few years should
help to significantly decrease these uncertainties and ascertain the
exact orbit of the binary. We note that our orbital solution differs
significantly from that obtained in the past, in the sense that we
find that the stars are close to apoastron while R96 concluded that
the stars must be close periastron. This contradiction is the result
of our estimated velocity being much smaller than the earlier
estimate, and smaller than the velocity one would observed if the
orbit was circular. A similar conclusion, based on velocity comparison
alone, was recently reached by Krist et al. (2002). Our derived orbit
is shown in Figure~\ref{orb}, where we have also plotted yearly
weighted averages of the data points used in the fit.

In the framework of investigating interactions between the ring and
the inner binary, it is interesting to compare the inner radius of the
circumbinary disk to the semi-major axis of the system. The ratio of
these two quantities lies in a 1$\,\sigma$ range $3.2 < R_{in}/a < 6.7$;
the two stars have cleared a wide gap around them. Artymowicz \& Lubow
(1994) have used analytical and numerical approaches to study the
formation of such a gap through gravitational resonances in
geometrically thin disks around circular and eccentric binary
systems. The orbital properties of GG Tau A provide us with the first
direct test of these models in pre--main-sequence visual binary
systems, although uncertainties prevent us from reaching definitive
conclusions. Artymowicz \& Lubow found that this gap clearing
phenomenon is primarily driven by the orbital eccentricity, with a
smaller dependence on the mass ratio, ($M_{B}/M_{A}\sim0.9$ in the case of GG
Tau A; White et al. 1999). The results of Artymowicz \& Lubow can be
summarized as follows: while the $R_{in}/a$ ratio is on the order of
1.7 for circular orbits, it can grow up to $\sim3.3$ for highly
eccentric binaries ($e=0.75$). The derived orbital parameters appear
somewhat problematic, with a $R_{in}/a$ ratio that seems to be larger
than those expected for a moderately eccentric system. This may mean
that the dynamical evolution of the GG Tau circumbinary ring is more
complex than previously thought.

\subsection{Circumbinary Disk Properties}
\label{cbdisk}
\subsubsection{Disk Detection \& Morphology}
\label{1-D}
Azimuthal averages of the PSF subtracted images (see
Figure~\ref{profile}) show that the circumbinary disk is easily
detected at radii greater than $\sim$1\arcsec\ at all three
wavelengths.  The circumbinary disk is detected with an average SNR
$>$ 5 per pixel at radii of 0\farcs9 to $\sim$1\farcs9\ (F110W and
F160W) and 1\farcs1 to 1\farcs9 (F205W data) with an average peak
surface brightness of 15.3, 14.1 and 14 mag/arcsec$^{2}$ in F110W,
F160W and F205W respectively. At smaller radii the 1$\sigma$
subtraction noise level rises dramatically, from $\sim$14
mag/arcsec$^{2}$ at 0\farcs5 to $\sim$11 mag/arcsec$^{2}$ at 0\farcs2,
with each filter having a slightly different noise level (see
Figure~\ref{profile}). This prevents a meaningful investigation of
gap material that has been posited in varying forms in earlier 
observations (R96; Silber et al. 2000; Krist et al. 2002).

A model of the two-dimensional near-infrared apparent disk geometry is
constructed from the F110W and F160W data by analyzing the intensity
profile of the disk as a function of azimuth.  Note that what we are
observing here is the geometry of the optically thick scattered light
distribution, not the true disk geometry. For every 10 degree segment
of the disk, an average radial profile is produced. The position of
the radial profile peak value is found using two methods.  The first
calculates a weighted centroid over a 0\farcs5 region centered on the
maximum value of the radial profile, which works well on sections of
the disk that have a distinct peak. The south side of the disk, however, has a
much flatter radial profile, and the weighted centroid method is not
as stable for these segments. The second method fits the radial
profile with a 4 degree polynomial. The `center' of the disk is then
calculated by finding the midpoint between the two radii at which the
disk value is equal to half the peak value.  For the north side of the
disk where the radial profile is almost gaussian in shape, this method
finds the same disk peak. For the flatter southern profiles it
provides a more robust method of finding the disk center. The final
peak position values are taken from the results of the polynomial
fit.  This method of finding the disk peak location also provides an
estimate of the width of the radial profile, which is taken to be the
distance between the two half-peak radii.  Uncertainties for these
values are estimated by taking the largest of either the difference
between the values found through the two methods or the standard
deviation of the values found from applying the same analysis to 3
subsets of the data.

The overall apparent shape of the disk is explored by fitting an
ellipse to the positions of the radial profile peak values.  We find
an ellipse with a semi-major axis of 1\farcs42 $\pm$ 0\farcs06 (200
AU) orientated at a position angle, $PA_{NIR}$ of 21$^{\circ}$ $\pm$
9$^{\circ}$ and an eccentricity of 0.64 $\pm$ 0.02. Although only G99
provided uncertainties for their analysis, our measurements of the
ring geometry are consistent with that found in both the previous
near-infrared (R96) and millimeter observations (Dutrey et al. 1994;
G99). If you assume that the disk is intrinsically circular and
geometrically thin, the observed eccentricity corresponds to a disk
inclination of 40$^{\circ} \pm 2^{\circ}$. While this is consistent
with inclination measurements made from the optically thin millimeter
images of the circumbinary ring ($i=37^{\circ} \pm 1^{\circ}$,
$PA_{disk}=7^{\circ} \pm 2^{\circ}$; G99), the disk is known to be
geometrically thick, which should bias the observed inclination
towards slightly larger values.  G99 also note that the geometrical
thickness of the inclined ring will cause the northern inner edge of
the scattered light ring to appear closer to the center of mass of the
system than the inner edge observed in the optically thin millimeter
images. This offset can be used to calculate the total height of the
disk above the midplane, i.e., where the NIR light becomes optically
thin. This height is where the disk is physically truncated in the
Monte Carlo simulations (see \S\ref{discus}).  In the NICMOS images,
the northern inner edge (the peak of the radial profile) occurs
$0\farcs88 \pm 0\farcs07$ away from the center of mass. From the disk
model fits to the millimeter images (G99), the inner edge of the
millimeter disk occurs at a projected distance of $R\cos i = 1\farcs03
\pm 0\farcs02$. The offset, $0\farcs15\pm0\farcs07$, corresponds to a
total disk height of $35^{+21}_{-18}$\,AU (using the relation from
G99). This is smaller than the 0\farcs25 offset derived by G99 from
the deconvolved adaptive optics images of R96.  In addition, the disk
height will also translate into an offset between the center of the
ring and the center of mass of the system. The disk fitting routine
finds that the center of the ring is offset from the center of mass of
the binary by 0\farcs21 $\pm$ 0\farcs03 along a position angle of
$162^{\circ} \pm 9^{\circ}$ degrees, consistent with the offset
measured from the inner edge.

The width of the disk has a clear azimuthal dependence, with the North
side being narrower than the South side, as noted by both Silber et al. (2000)
and Krist et al. (2002). Figure~\ref{width} shows
the measured width, which ranges from 0\farcs3 to $\sim$1\arcsec, as a
function of position angle. We fit these measurements with a modified
Henyey-Greenstein scattering phase function (described in
$\S$\ref{azvar}), which is physically meaningless in this context but
provides an analytical description of the data. This analysis shows
that the thinnest portion of the disk occurs at a position angle of
356$^{\circ}$, independent of wavelength (see Table \ref{azfit}). Because
the function we used for the fit is unphysical and provides
only a moderately good fit, we do not estimate uncertainties.

At the lowest contour levels we observe a deviation away from the
elliptical fit along the southern edge of the disk (see
Figure~\ref{kink}). This deviation consists of a sharp elbow, or
kink, in the south-east edge of the disk and a straightening of the
isophotes along the southern edge which makes the disk outer edge
appear boxy. This effect is present in all 3 filters, and coincides
with the kink detected by Silber et al. (2000) and noted by Krist et al. (2002).

\subsubsection{Integrated Disk Photometry}
\label{colors}
In order to assign flux densities to the disk, we define an annular
aperture with an eccentricity of 0.64 centered at the position of the
center of the ellipse with an inner semi-major axis at 1\arcsec\ and
outer semi-major axis at 2\farcs2, based on the results given in
\S\ref{1-D}.  The area affected by the diffraction spikes is
particularly noisy and is excluded from the aperture by masking off
these regions, reducing the effective aperture area from 9.2 to 5.82
square arcseconds. The spike mask comprises two diagonal stripes,
0\farcs8 in width, centered on the position of the primary. Summing
the counts over this area and correcting the flux for the masked off
regions (by multiplying the area of the masked regions with the
mean disk flux per pixel), provides the estimates of the total disk magnitudes
given in Table \ref{astrom}.  Since the pixel scale is smaller than the
diffraction limit, the estimated uncertainties are based on maps that
have been averaged over regions corresponding to the diffraction
limited beam sizes (0\farcs1, 0\farcs16 and 0\farcs2 for F110W, F160W
and F205W respectively). These measurements of the integrated disk
intensity have an order of magnitude higher signal to noise ratios
than earlier ground-based measurements.

The overall disk magnitudes are compared to those derived for the
binary system (\S\ref{orbit}) to obtain disk to star flux ratios and
color excesses (see Table~\ref{nearfar}). The NICMOS images presented
here lead to a disk to star flux ratio of $\sim$1.5\%, which is 2.5
times larger than that derived by R96. Likewise, the color of the
circumbinary disk is comparable to that of the stars: 

\begin{center}
$\Delta(M_{F110W}-M_{F160W}) = 0.10 \pm 0.03 $\\
$\Delta(M_{F160W}-M_{F205W}) = -0.04 \pm 0.06 $
\end{center}

\noindent where $\Delta(M_{1}-M_{2})$ is the color excess of the disk, 
or $(M_{1}-M_{2})_{disk} - (M_{1}-M_{2})_{star}$. This is in 
contrast to the large red excess suggested earlier.

\subsubsection{Spatially Resolved Disk Photometry}
\label{azvar}
The spatially resolved color properties of the circumbinary disk are
investigated by calculating the disk magnitudes within circular
apertures which are 0\farcs19 in diameter (roughly the size of the 2
\micron\ diffraction limit) placed around the disk. For this
comparative analysis, the F110W and F160W bias-subtracted images have
been convolved to the resolution of the F205W image. Only areas of the
disk used to calculate the integrated disk photometry are included in
this analysis. Figure~\ref{diskmag} displays the resulting
color-magnitude plot, which shows no significant trend in color with
respect to magnitude. Unlike the results from R96, the NICMOS
measurements provide no evidence for extinction within the disk.

As originally mentioned by R96, the disk exhibits azimuthal variations
in intensity that are presumed to arise from angular variations in the
scattering efficiency, with the brightest side of the ring
corresponding to forward scattering from the edge of the disk that is
nearest to our line of sight. Here we present a detailed quantitative
analysis of these azimuthal variations, using normalized peak flux
densities derived from the 10$^{\circ}$ azimuthal averages discussed
in \S\ref{1-D}. To estimate the position angle of the brightest
region, the data are fit, using a chi-square minimization technique,
with a modified\footnote{The following two small modifications have
been made to the Henyey-Greenstein scattering phase function for
fitting the position angle of the brightest region: (1) the albedo,
which normally is included as a multiplicative factor, is omitted,
since we have normalized the data and (2) the position angle of the
brightest region, $PA_{0}$, is added.} form of the Henyey-Greenstein
scattering phase function (Henyey \& Greenstein, 1941)

\begin{displaymath} 
S(PA) \propto [1-g'^{2}][1+g'^{2} - 2g'cos(PA-PA_{0})]^{-3/2}
\end{displaymath}

\noindent where $PA_{0}$ is the value of position angle that maximizes
the function. While the use of this function is not physically
meaningful, as the phase function depends on scattering angle, not
position angle on the sky, it does provide a convenient method of
characterization. The best-fit functions are displayed with the data
in the left hand column of Figure~\ref{phase} while the values of
$PA_{0}$ and the fitted peak-to-peak near side to far side flux ratios
are listed in Table~\ref{azfit}. The peak near side to far side flux
ratio is $\sim$4 with no significant wavelength dependence.  The
position angle of the brightest portion of the disk ($22^{\circ} \pm
4^{\circ}$) is roughly consistent with that measured by Krist et
al. (2002) in I band images (25$^{\circ}$) and both the position angle
of the semi-minor axis ($7^{\circ} \pm 2^{\circ}$, G99) and the
position angle of the thinnest region ($-4^{\circ}$, derived in
\S\ref{1-D}). This provides strong support for the hypothesis that the
disk is geometrically thick.

The azimuthal variations in intensity arise from variations in dust
grain scattering efficiencies as a function of scattering angle,
rather than the observed position angle around the disk. Assuming a
single scattering scenario in a geometrically thin disk, Close et
al. (1998) derive scattering properties of the dust grain population
by matching the intensity extrema to scattering angles of $90-i$ and
$90+i$, which, with an assumed inclination, $i$, of 35$^{\circ}$,
results in scattering angles of 55$^{\circ}$ and 125$^{\circ}$. The
NICMOS images, however, contain much more information than just the
intensity extrema, as the circumbinary disk samples a large range of
scattering angles as a function of position angle around the disk.  We
therefore take the additional step of fitting the entire azimuthal
variation of flux with a scattering phase function.  Using the known
disk geometry, the observed variation with position angle can be
translated, or de-projected, into a variation with scattering angle,
$\theta$, which is defined as the angle of deflection away from the
forward direction of the incoming light.  $\theta$ is geometrically
related to the position angle around the disk, $PA$, the position
angle which maximizes the azimuthal variation, $PA_{0}$, and the disk
inclination, $i$, by:

\begin{displaymath}
 cos(\theta+\phi_{open}) = \sqrt{1 - \frac{1}{1+cos^{2}(PA-PA_{0})tan^{2}(i)}} \times (-1)^{j} 
\end{displaymath}
\noindent where
\begin{center}
$j = 1$ if $cos(PA) < 0$ \\
$j = 0$ if $cos(PA) > 0$
\end{center}   

\noindent A derivation of this relation can be found in Appendix
~\ref{getscatang}.  $\phi_{open}$ is the opening angle of the disk as
seen from the stars; including this factor converts the relation from
a geometrically thin disk to a thick disk case. It is assumed
here that, at all position angles on the sky, the intensity is
dominated by the portion of the ring corresponding to the smallest
scattering angle, which is true for forward-scattering 
grains. Using the values for the inner radius of the disk and a
disk height of 35\,AU, we calculate $\phi_{open}$ to be 11$^{\circ}$.
In this case, with an inclination of 37$^{\circ}$, the
scattering angle ranges from $\sim$40$^{\circ}$ at the closest edge
to $\sim$120$^{\circ}$ at the edge furthest from us.  Using the
above relationship to translate the observed position angle into
scattering angles, the intensity variations are then fitted with a
normalized Henyey-Greenstein scattering phase function, in order to
estimate the value of the dust asymmetry parameter, $g$. This parameter is the averaged
cosine of the scattering angle which ranges in value from -1 to 1. 
Given that we are using the one parameter Henyey-Greenstein scattering phase function
we can only investigate forward throwing dust grains with $0 < g < 1$. In
this case, if the
light is scattered isotropically, then the amount of forward scattered
light equals that being back-scattered and $g$=0.  As the light becomes
more strongly forward-scattered, the value of $g$ increases.
The results of the fit for the dust asymmetry are summarized in
Table~\ref{azfit}. We find that omitting the disk opening angle 
will slightly overestimate $g$. It should be recalled 
that these values are obtained under the assumption of single scattering
(see \S\ref{multiplescat} for further discussion).

Since the system may be modulated by geometric factors, such 
as a possible shadowing by circumstellar disks, Wood et al. (1999) 
found it more convenient to work with an integrated
quantity to describe the near side to far side flux ratio.
Integrating the flux over the northern and southern halves of the
disk, as defined by the position angle of the disk, we find the ratio
of scattered light to be $\sim$1.4, with no significant wavelength
dependence (see Table \ref{nearfar}). This value is roughly a factor
of two smaller than that seen in the ground-based AO images (Wood et
al. 1999).

\section{Discussion}
\label{discus}
In this section, we use the various quantities derived in \S\ref{res}
to revisit the question of whether the existence of large dust grains
or opaque circumstellar disks are required to explain the observed
color excesses. Monte Carlo scattering simulations are presented,
qualitatively reproducing the NICMOS data (\S\ref{MCmodel}) and 
emphasizing the key role played by grain properties, especially the dust asymmetry 
(\S\ref{multiplescat}).  The scattering simulations show
that neither circumstellar disks or large grains are absolutely needed
to reproduce the data and that our fitted scattering phase function
can be used to derive a lower limit on the median size of the
scatterers.

\subsection{Numerical Modeling of the GG Tau Ring}
\label{MCmodel}

Since the NICMOS measurements of the GG Tau
circumbinary disk provide different results from both those previously
reported and modeled, we ran new Monte Carlo simulations to
investigate whether the observed disk colors can be reproduced using a
standard ISM grain model.  We began, like Wood et al. (1999), with the
simplest case: modeling the scattered light distribution from a thick
torus of ISM-like dust, without the additional complications of either
circumstellar disks or grain growth (their model 1).

We used a Monte Carlo code that was readily available to us; the same
model which successfully reproduced the NICMOS polarization results
(Silber et al. 2000). The Monte Carlo code is based on a program from
M\'enard (1990) that has been modified such that each photon interacts
with a randomly sampled dust grain from the input grain size
distribution.  Details of the modifications can be found in Duch\^ene
(1999). The disk is modeled as a thick torus, using a total disk
height of 38 AU at the inner edge of the ring, close to that estimated
from the observed inner edge offset (see \S\ref{1-D}).  This torus has
a total disk mass of 0.13 $M_{\odot}$ and a scale height of 21\,AU at
the inner edge of the ring (180\,AU) that varies as $r^{1.05}$
(G99). Outcoming photons are sorted by inclination: all results
presented here are for the range of inclinations closest to the actual
inclination of the system, $37^{\circ} < i < 46^{\circ}$. The
simulations follow 1 million photons per filter. The dust grain
properties are taken from the Mathis \& Whiffen (1989, hereafter known
as MW) dust model. While the dust parameters (e.g., albedo and dust
asymmetry) in the MW dust model differ significantly from the Kim,
Martin \& Hendry (1994; hereafter known as KMH) model\footnote{MW
dust grains are composite and porous in nature, composed of silicate,
graphite, amorphous carbon and vacuum. The grain size distribution
follows $n(a) \propto a^{-3.7}$ between 0.03 and 0.9 \micron\,(MW
model A for an $R_{V}$ = 3.1). In contrast, the KMH dust grains are
composed of separate silicate and graphite particle populations, with
a size distribution of $n(a) \propto a^{-3.5}exp(-a/0.2\micron)$
between 0.005 and 1 \micron.} used by Wood et al. (1999), both models
successfully reproduce the ISM extinction curve. The simulated disk
properties can be found in Table~\ref{nearfar}. The resulting Monte
Carlo simulations (see model 1 in Table~\ref{nearfar}) roughly
recreate the observed disk to star flux ratios and near side to far
side flux ratios. They also produce a considerably red disk color
excess, $\Delta(M_{F110W} - M_{F160W})=0.50$, $\Delta(M_{F160W} -
M_{F205W})=0.28$. This is distinctly larger than the color excess
observed with NICMOS, and very different from the blue scattered light
disk seen in the Wood et al. (1999) models.

The two main differences between the Wood et al. simulations and those
presented here are the dust grain properties used\footnote{Note that there
is also a slight difference in the way the dust parameters are calculated. Wood et al.
calculate the mean dust parameters for the grain population and assign these
to each dust particle, whereas the
dust parameters for each photon-dust particle interaction in the simulations presented
here are calculated on the fly after the grain size has been randomly sampled from the grain 
size distribution. While this method is more computationally expensive, we believe it
to be more representative of the actual phenomenon. How large an effect this
difference makes has not been investigated so far.} and a slightly
different disk geometry.  Both the overall disk height and the degree
of flaring differ (Wood et al. use a scale height that varies as
$r^{1.25}$ from the earlier millimeter work of Dutrey et al. 1994). 
To fully compare our numerical results with those of Wood et al.,
we ran Monte Carlo simulations where the only difference is the dust
grain properties. The results from these simulations (see model 2
in Table~\ref{nearfar}) show that the red disk color excesses remain and
therefore are predominantly caused by the dust grain properties.

\subsection{The Influence of Multiple Scattering and Grain Properties: \\
Explaining the Different Modeled Colors}
\label{multiplescat}

The apparently contradictory simulations presented here and in Wood et
al. can be reconciled by investigating the effect the different dust
grain properties have on the scattering process.  In optically thick
situations, photons are scattered multiple times. The resulting
scattered light surface brightness is dependent on two key dust
parameters, the albedo of the dust, $\omega$, and the azimuthal scattering
dependence given by the dust asymmetry parameter, $g$. In terms of disk 
geometry, the inclination of the system is also an important factor. 
At any one scattering angle, the output
scattered surface brightness is dependent on both the albedo of the
dust and the number of scattering events $n$, such that $SB_{scat}
\propto \omega^{n(g_{\lambda})}$ (e.g., Witt \& Oshel, 1977; Witt 1985).  The greater
the number of scatterings, which is a function of the dust asymmetry
parameter, the greater reduction in the resulting scattered surface
brightness. 

The simulations presented here and those in Wood et al. (1999) are
based on the same Monte Carlo scheme and have the same disk
geometry. The main difference lies in grain properties used, and this
alone is enough to explain the disk color reversal between the two
sets of models. Table~\ref{dustprops} compares the dust properties used
in the two simulations. The dust asymmetry of the median scatterer in
our simulations is $\sim$2 times larger than that used by Wood et al.,
while the albedos are essentially the same. How does this affect the
observed scattered light distribution? Consider first the far side
(southern edge) of the ring where back-scattering is the
dominant effect. For low $g$ values, such as those used in the Wood et
al. simulations, the scattering is not far from being isotropic, which
means that the photons leaving the disk toward the line of sight will
in general have gone through a small number of scatterings ($\sim$ 1).
Thus the resulting surface brightness distribution
will not be strongly affected by multiple scattering effects and the
scattered light colors will be similar to the wavelength dependence
of the dust albedo, resulting in Wood et al.'s case in blue disk color
excesses. For more strongly forward scattering dust particles (large
$g$), back-scattering is rarer and in this region of the disk most of
the scattering events will cause photons to move further into the
disk, away from our line of sight. The photons that do escape from the
surface of the disk toward our line of sight will have undergone a
much larger number of scatterings and the resulting scattered light
surface brightness will be significantly changed by the wavelength
dependence of the dust asymmetry parameter. For the MW dust
properties, not only is $g$ larger than those in the KMH model, they also have 
a stronger wavelength dependence. The scattered light surface brightness will be
more reduced at 1 \micron\, than it is at 2 \micron\,
producing significantly red disk colors. The same story applies
to the nearest (northern) edge of the disk, where we are predominantly 
seeing forward scattering off the upper limb of the ring. 
This type of scattering is enabled by both low and high $g$ values (near-isotropic and strongly
forward scattering); while the number of scattering events remains small in
both simulations there will still be more scattering events for the
photons in our simulation than those in Wood et al.'s due to the
inclination of the system favoring low $g$ values slightly more.
This, combined with the corresponding slightly longer pathlength 
in our simulation, enabling more extinction, is
enough of an effect to compensate for the intrinsically blue color of the
scattering, as seen in our simulations. Taking the disk as a whole,
this explains why our simulations predict a red disk color, while those
of Wood et al., being dominated by the flux from the far side of the
disk, result in overall blue disk colors. The same
$\omega^{n(g_{\lambda})}$ dependency is responsible for the reduced
disk to star flux ratios and increased near to far flux ratios
observed in our simulations, as photons undergo more scattering events
in the far side of the disk than in Wood et al.'s (see Table~\ref{nearfar}).
Given the range of colors that can be simulated using the two
grain models, it could be expected that a grain population with properties
intermediate of those in the KMH and MW models would result in neutral near-infrared
colors, as observed.

An additional effect of multiple scattering observed in our
simulations is that the disk inclination and optical thickness combine
to modify the azimuthal intensity variation from that expected from
single scattering. Given that the observed surface brightness is
dominated by the least scattered photons ($\omega^n$ dependency), for
grains that are predominantly forward-scattering, photons received
from the back side of the ring are biased towards the back scattering
part of the phase function. Statistically, most of the photons will be
scattered towards the middle of the ring and will either not reach the
observer or be of an extremely low intensity. Scattering from the front
edge of the ring, however, will remain mostly unbiased. This biasing 
effect will cause the derived $g$ value measured in \S\ref{azvar} 
to underestimate the true dust asymmetry, and hence our value is
only a lower limit. The same reasoning applies to the
Close et al.'s analysis, as a consequence of considering only single
scattering. As an example of this effect, in our simulations, the size
of the median scattering grain is 0.56 \micron\ and these have an
asymmetry parameter $g\sim0.8$ at 1 \micron. However, the measured $g$
from the simulations (obtained using the same method as in
\S\ref{azvar}) is only $\sim0.3$.

The observed lower limit on g, measured in \S\ref{azvar} provides a
lower limit on the median size of the scatterers as the asymmetry
value is a sensitive function of the size parameter, $x$ ($=2\pi a /
\lambda$), for $g$ less than $\sim$0.8. From the MW grain properties used, 
we find that the measured asymmetry of $g > 0.39$ at 1 \micron\ corresponds to
a grain size, $a >$ 0.23 \micron.

\section{Implications and Summary}
\label{implications}

The GG Tau disk is clearly detected ($SNR_{pix} > 5$) in all three
NICMOS filters between radii of 1\farcs0 to 2\farcs2, scattering
approximately 1.5\% of the stellar light.  No discernible wavelength
dependence of scattered light is observed, although azimuthal
variations in both width and intensity are seen and modeled.  We
confirm the observation of a kink in the disk by Silber et al. (2000).
Observed as a sharp elbow in the disk on the south-east edge, followed
by a flattening of the isophotes along the southern edge of a disk,
this feature is seen in all three filters. The projected distance of the 
northern edge from the center of mass, related to the total height of the ring,
provides the first hint that the ring may not be as thick as previously estimated, 
with a total height of 35\,AU at the inner edge of the disk (180\,AU).

The presence of such a well-observed circumbinary ring around this
system, combined with the 9 year baseline of binary observations has 
allowed us to constrain the orbital parameters of the
binary.  We find a slightly elliptical orbit $e=0.3 \pm 0.2$, with a 
semi-major axis $a=35^{+22}_{-8}$\,AU and a corresponding orbital period
of $185^{+195}_{-55}$\,yr. The 1$\sigma$ range of estimated orbital semi-major
axis is such that the ratio of semi-major axis to inner disk radius is
larger than expected from previous SPH simulations (Artymowicz \& Lubow, 1994).
This suggests that the dynamical evolution of the circumbinary ring 
may be more complex than previously thought. Further astrometric
observations of this close binary are needed to constrain the binary
orbit further.

The Monte Carlo simulations summarized in \S\ref{MCmodel} show the
wide range of colors one can `predict' from the GG Tau disk, from blue
to red, using different dust size distributions and grain properties
($\omega$, $g$). Specifically, our Monte Carlo simulations show that
the observed neutral disk colors, ratio of disk to star light, and
near side to far side flux ratio can probably be reproduced with some
combination of an ISM-like grain size distribution and dust properties
and does not necessarily require either the presence of large dust
grains or prior extinction by circumstellar disks. However, this does
not rule out that either is present. Because the disk is optically
thick in the near-infrared, studies of scattered light are only
sensitive to those grains that are located near the surface of the
ring. Any conclusion about grain growth only applies to those regions
and does not provide any information the size of grains deep in the
ring.  For instance, it has been shown that grains larger than $\sim
1\,\micron$ tend to segregate in the disk midplane because of their
larger mass (e.g., Suttner \& Yorke 2001) and possibly as the result
of the enhanced grain-grain collision rate in higher density
regions. The presence of such large grains in the midplane cannot be
determined in our near-infrared observations. Circumstellar disks are
known to surround both components of GG Tau A, however, the possible
effects they have on the integrated circumbinary disk colors have been
shown to be subtle and somewhat surprising.  While Krist et al. (2002)
suggest that the red disk colors observed in the optical could be
caused by an $A_{V} \geq$ 1.2 mag of extinction, possibly in the form
of circumstellar disks, Wood et al. (1999) find that the inclusion of
small circumstellar disks can actually cause the integrated disk
colors to become {\it bluer} rather than redder. Analysis of the 
disk colors at both optical and near-infrared wavelengths 
is planned for future work. 

While the color of scattered light in optically thick disks cannot
currently be used to constrain grain properties, the azimuthal
intensity variations can be used to provide a lower limit on the
median grain size, independent of grain model. This lower limit ($g >
0.39$ at 1 \micron\,) corresponds to a median grain size of $a > 0.23$
\micron.  Given that the ISM dust models produce a wide range of
median grain sizes ranging from $\sim$0.16 \micron\,(Mathis, Rumpl \&
Nordsieck 1977; KMH) to 0.56 \micron\,(MW), this lower limit 
does not suggest that grain growth has occured in the surface layers of 
the disk in this young ($\sim$1 Myr old; White et al. 1999) system.

There is still considerable uncertainty regarding the ISM dust
properties (e.g., Witt 2000), a point which is well illustrated here
by the comparison of Monte Carlo scattering results using two of the
more well known dust grain models. Both the KMH and MW grain models,
like most of the ISM dust models to date, have been constrained using
the ISM extinction curve, which they reproduce equally well.  The
results presented here suggest that using the extinction curve {\it
alone} to constrain the dust properties introduces a significant level
of uncertainty in the scattering properties of the grains. As an
example, KMH and MW grain models predict V band $g$'s of $\sim0.5$ and
$\sim0.9$ respectively. Observations of reflection nebulae in the same
bandpass infer $g\sim0.7$, significantly different from either of
these models (Witt, Oliveri \& Schild, 1990).  Additionally, models of
T Tauri disks seen in scattered light with WFPC2 find that the
observations are best fit by $g\sim0.65$ (Burrows et al. 1996;
Stapelfeldt et al. 1998; Krist et al. 2002), although these values may
be affected by grain growth.  Combining additional sources of
information, such as observations of dust scattering (e.g., Witt, et
al. 1990), polarization (e.g., Zubko \& Laor 2000), and dust thermal
emission (e.g., Li \& Draine 2001) will provide additional
constraints. Without a more detailed knowledge of the ISM dust grain
properties it is unlikely that we will be able to unambiguously
determine whether grain growth is occuring in T Tauri disks through
scattered light imaging.

Although dust properties in the disk cannot currently be constrained
from either colors or dust asymmetry alone, this work provides an
outline for future analysis and modeling. Observationally, further
resolved intensity and polarization maps at other wavelengths are key
in sampling a large enough grain size parameter ($x$) range to
constrain the dust grain properties. For instance, at 3 to 5 \micron\,
the dust asymmetry is expected to be much more isotropic, allowing
significant color variations (e.g., $J-L$) between the front and back
sides, unless significant grain growth has already occurred. Such
observations should be carried out in parallel with a full exploration
of the parameter space (grain size distribution, grain properties,
ring geometry) through Monte Carlo scattering simulations.

The authors thank Mike Jura, Alycia Weinberger, Fran\c{c}ois M\'enard, 
Kenny Woods, Lisa Prato, and
Angelle Tanner for enlightening discussions.  We also thank the anonymous
referee for their constructive comments. Support
for this work was provided by NASA through grant number
GO-06735.01-95A from the Space Telescope Institute, the NASA
AstroBiology Institute and the Packard Foundation.  This research was
based on observations made with the NASA/ESA Hubble Space Telescope,
obtained at the Space Telescope Science Institute, which is operated
by the Association of Universities for Research in Astronomy, Inc,
under NASA contract NAS5-26555. This research has made use of the
SIMBAD database, operated at CDS, Strasbourg, France.

\appendix
\section{Determination of the orbital parameters of the binary}
\label{getorb}

In this appendix, we present the method used to derive the orbital
parameters of the binary from a single measurement of its separation
and relative velocity with some additional knowledge available from
previous studies.

The orbit of a binary system can be described with a set of 9
independent parameters: mass ratio ($q$), eccentricity ($e$), total
mass ($M$), orbital period ($P$), time of periastron ($T_0$),
inclination of the orbital plane ($i$), position angle on the sky of
the ascending node of the orbit ($\Omega$), angle in the orbital
plane between periastron and the ascending node ($\omega$), and distance
to the system ($D$). Note also
that $P$ and $M$ are linked to the semi-major axis of the orbit, $a$,
through Kepler's third law. When considering the {\it relative} motion
of one component around the other one, $q$ is no longer a relevant
parameter. Therefore, one must measure 8 independent quantities to
solve the orbit unambiguously. In the case of GG\,Tau, the millimeter
interferometric measurement of the circumbinary ring by G99 yields
precise estimates of three independent parameters: the inclination of
the ring, $i'=37\degr\pm1\degr$, the position angle of the apparent
semi-minor axis of the ring, $\Omega'=7\degr\pm2\degr$, and the total
system mass $M=1.28\pm0.07\times (D/140\,{\rm pc}) M_\odot$. If the
ring is intrinsically circular then $\Omega=\Omega'\pm90\degr$
(depending on the motion of the binary). Furthermore, if we assume
that the orbital plane corresponds to the ring mid-plane, we have
$i=i'$. These two assumptions (circular ring and coplanar orbit) are
the only assumptions we make in the following analysis. Additionally,
the HIPPARCOS determination of the distance to the Taurus star-forming
region, $D=139\pm10$\,pc, by Bertout et al. (1999), allows us to
convert the apparent separation and velocity of the binary into
absolute values. Prior to our work, the problem could thus be
simplified to a problem with only four unknowns.

The orbit of the system has been followed over the last ten years or
so and, since all measurements so far are consistent with a uniform
linear motion, we combined them all to obtain the average binary
separation and linear velocity projected on the sky. These quantities
can be represented for instance by the respective amplitude of the
binary separation and of its velocity ($\rho$ and $\dot{\rho}$,
measured in arcsec and arcsec.yr$^{-1}$) and by the angle between
these two vectors in the plane of the sky ($\delta$). In the following
derivation, a useful quantity is the angle $\theta$ between the 
apparent semi-minor axis of the ring and the binary separation, which represents
the polar coordinate of the current measurement with respect to the
disk itself. This angle can be expressed as
$\theta=\Omega'-PA_{bin}$. Our data imply
$\rho=0\farcs2485\pm0\farcs0013$,
$\dot{\rho}=(6.68\pm0.52)\,10^{-3}\,\,\arcsec{\rm yr}^{-1}$,
$\delta=96\fdg0\pm5\fdg5$ and $\theta=11\fdg2\pm2\fdg1$ (see
\S\ref{orbit}). The orientation of the relative velocity implies that
$\Omega=97\degr$. We have thus obtained the four last measurements
needed to solve the whole orbit.

The two-body problem is a classical mechanics problem, the solution of
which can be found in many textbooks. We are interested here in the
case where only relative measurements of the 3D binary separation
($\overrightarrow{r}$) and velocity ($\overrightarrow{v}$) are
available from the de-projection of the observed separation and motion
in the orbital plane. The eccentricity and semi-major axis can then be
derived from measurements of the angular momentum,
$j=|\overrightarrow{r}\times\overrightarrow{v}|$, and total energy,
$E=\frac{1}{2}v^2-\frac{GM}{r}$, {\it per unit mass}. We have the
following relations: $e^2=1+\frac{2j^2E}{G^2M^2}$ and
$a=-\frac{GM}{2E}$. These relations can be translated into the
observables presented above as follows:
$$
e^2=1+\frac{\rho^2{\dot{\rho}}^2D^4\sin^2\delta}{G^2M^2\cos^2i}\left\{
{\dot{\rho}}^2D^2
\left[\sin^2(\delta+\theta)+\frac{\cos^2(\delta+\theta)}{\cos^2i}\right]
-2\frac{GM\cos i}{\rho D} \right\} $$ and $$ a=GM\left\{2\frac{GM\cos
i}{\rho D}-{\dot{\rho}}^2D^2\left[
\sin^2(\delta+\theta)+\frac{\cos^2(\delta+\theta)}{cos^2 i}\right]
\right\}^{-1}$$

One can immediately derive the orbital period: $P({\rm yr}) =
\sqrt{\frac{a^3({\rm AU})}{M(M_\odot)}}$. Uncertainties for all of
these quantities were obtained by allowing all seven observables to
vary within 1$\,\sigma$ of their nominal value and retaining the
extreme values for each orbital parameter. Special care must be taken
when considering the uncertainties induced by the distance estimate,
as the total mass of the system is also affected in a {\it systematic}
manner by the distance. The distance and relative velocity of the
binary estimates combine for the largest part of the total uncertainty
for the orbital elements.

For completeness, we derive the remaining two orbital parameters that
have not been addressed so far, $\omega$ and $T_0$. The former is the
angle, in the orbital plane, between periastron and the ascending
node. This can be derived from the angular distance between the
current location of the companion and periastron. The latter angle can
be determined using the general equation of the elliptical orbit,
$$r=\frac{a(1-e^2)}{1+e\cos (\phi-\varpi)}$$ where $r$ and $\phi$ are
the polar coordinates of a point running along the orbit and $\varpi$
is the position angle of periastron. Using the current de-projected
separation of the system, 43.0\,AU, we find that the binary is about
$(\phi-\varpi)=-145\degr$ away from periastron\footnote{The negative
sign reflects the fact that the binary is slowly getting tighter,
hence closer to periastron.}. Combining this information with the
angle between the current binary location and the ascending node
($100\degr$), we conclude that $\omega=245\degr$. To estimate how much
time it takes for the companion to go from one point to another along
the orbit, we use Kepler's second law and numerically integrate the
area $A$ defined by the section of an ellipse between these two
points. We can easily calculate the time derivative of the area, which
is related to the angular momentum per unit mass by $\frac{{\rm
d}A}{{\rm d}t}=\frac{1}{2}j$. Applying this to the average position of
the companion and periastron, we find that the next periastron passage
will occur about 61\,yrs from now or, equivalently, that
$T_0=2472467$\,JD. Both $\omega$ and $T_0$ depend heavily on the
previously derived quantities, therefore, we emphasize that the
uncertainties on these parameters are rather large.

\section{Calculating Scattering Angles}
\label{getscatang}
In this section we derive a relation between position angle around a
disk and the angle a photon is scattered through. At each position
angle around the disk, we assume that the observed photons are
scattered off the closest disk edge which corresponds to the smallest
available scattering angle (see \S\ref{azvar}).

Consider a circular ring that is inclined to the line of sight at an
angle $i$.  This ring is projected on the sky as an ellipse (see
Figure~\ref{model}a), with normalized semi-major axis, $d=1$ and
aspect ratio $c/d = cos i$.  The position angle on the sky of a random
point around the disk, $PA$, is defined as $tan(PA) = x/y$. In 
Figure~\ref{model}a we have rotated the image so that the semi-minor axis of
the disk is now vertical.  The amount of rotation, $PA_{0}$, is the
$PA$ of the semi-minor axis of the disk, defined in the usual manner.

The scattering angle is defined as the angle between the pre- and
post- scattering directions, $\vec{n}$ and $\vec{m}$
respectively, as shown in Figure~\ref{model}b.  This angle,
$\theta_{scat}$, is $\theta - \phi_{open}$, where $\phi_{open}$ is the
angle subtended by the disk height.  If the disk can be assumed to be
flat, we have $\theta_{scat} = \theta$. We first consider this case
before generalizing our results to $\phi_{open} \neq 0$.  The
coordinates of the unit vector $\vec{n}$, $(u,v,w)$ are along the
photon's initial path to a random point on the ring, $M (x,y)$, in the
$(\vec{x},\vec{y},\vec{z})$ frame are:

\begin{eqnarray}
u = x \\
v = y \\
w = \sqrt{(1-x^2 -y^2)} \times sign(cos(PA))
\end{eqnarray}

Writing the standard equation of an ellipse: 
\begin{displaymath}
(\frac{x}{d})^{2} + (\frac{y}{c})^{2} = 1
\end{displaymath}

\noindent and introducing the definitions of $i$ and $PA$, we find:
\begin{displaymath}
x^{2} \left( 1 + \frac{1}{tan^{2}(PA)cos^{2}(i)} \right) = 1
\end{displaymath} 

By definition, the dot product of the unit vectors, $\vec{n}$ and $\vec{m}$
is:
\begin{displaymath}
\vec{n}\cdot\vec{m} = \left( \begin{array}{c} 0 \\ 0 \\ 1 \end{array}\right) \left(\begin{array}{c} u \\ v \\ w \end{array}\right) = cos(\theta)
\end{displaymath}

\begin{eqnarray}
cos(\theta) = \sqrt{(1-x^2 -y^2)} \times sign(cos(PA)) \\
cos(\theta) = \sqrt{1-x^2(1 + \frac{1}{tan^2(PA)})} \times sign(cos(PA))\\
cos^2(\theta) = 1 - \frac{x^2}{sin^2(PA)} \\
\end{eqnarray}

Substituting the expression for $x^2$:
\begin{eqnarray}
1-cos^2(\theta) = \frac{1}{sin^2(PA)(1 + \frac{1}{tan^{2}(PA)cos^2{i}} ))} \\
                = \frac{1}{1 + cos^2(PA)\left(\frac{1}{cos^2(i)} -1 \right)} \\
                = \frac{1}{1 + cos^2(PA)tan^2(i)} \\
cos(\theta) = \sqrt{ 1 - \frac{1}{1 + cos^2(PA) tan^2(i)}} \times sign(cos(PA))
\end{eqnarray}

or
\begin{displaymath}
cos(\theta) = \sqrt{ 1 - \frac{1}{1 + cos^2(PA) tan^2(i)}} \times (-1)^j
\end{displaymath}
where
\begin{center}
$j = 1$ if $cos(PA) < 0$ \\
$j = 0$ if $cos(PA) > 0$
\end{center}

Reintroducing the position angle on the sky of the semi-minor axis of
the ellipse, $PA_{0}$ and the disk opening angle, $\phi_{open}$, we
can generalize this as:

\begin{displaymath}
cos(\theta) = cos(\theta_{scat}+\phi_{open}) = \sqrt{ 1 - \frac{1}{1 + cos^2(PA-PA_{0}) tan^2(i)}} \times (-1)^j
\end{displaymath}

For a face-on disk ($i =0^{\circ}$) the scattering angle is always
90$^{\circ} - \phi_{open}$.  Note that this relation between
scattering angle and disk inclination breaks down for edge-on disks
($i = 90^{\circ}$).

\begin{deluxetable}{cllcccc}
\tabletypesize{\small}
\tablecaption{Stellar and Integrated Circumbinary Disk Results\tablenotemark{a} \tablenotemark{b} \label{astrom}}
\tablehead{
\colhead{Filter} &\colhead{Separation} &\colhead{Position Angle}
&\colhead{Aa} &\colhead{Ab} &\colhead{Aa + Ab\tablenotemark{c}} &\colhead{Disk\tablenotemark{d}} \\
\colhead{} &\colhead{(arcsecs)} &\colhead{(degrees)} &\colhead{(mag)} 
&\colhead{(mag)} &\colhead{(mag)} &\colhead{(mag)}
}
\startdata
F110W &0.248 $\pm$ 0.002 &353.9 $\pm$ 0.4 &9.37 $\pm$ 0.05 &10.27 $\pm$ 0.07 &8.98 $\pm$ 0.02 &13.59 $\pm$ 0.01\\
F160W &0.251 $\pm$ 0.002 &353.8 $\pm$ 0.3  &8.39 $\pm$ 0.01 &9.08 $\pm$ 0.04 &7.91 $\pm$ 0.02 &12.42 $\pm$ 0.01\\
F205W &0.257 $\pm$ 0.003 &353.6 $\pm$ 0.8  &7.94 $\pm$ 0.01 &8.61 $\pm$ 0.01 &7.46 $\pm$ 0.02 &12.01 $\pm$ 0.05 \\
\enddata
\tablenotetext{a}{The zeropoints used are 1775, 1083 and 731 Jy in F110W, F160W and F205W respectively (M. Rieke, priv. comm).}
\tablenotetext{b}{The reported photometric uncertainties do not include the 
5\% uncertainty in the zeropoints as the analysis presented here
rests on the relative photometry of the disk compared to the central stars.}
\tablenotetext{c}{Combined binary magnitude is measured using wide aperture photometry and assuming
the disk is a small ($\sim1\%$) contamination on this.}
\tablenotetext{d}{Estimated disk brightness integrated over 9.2 arcsec$^{2}$ (see \S\ref{colors} for details).}
\end{deluxetable}

\clearpage

\begin{deluxetable}{lccccc}
\tablewidth{5in}
\tablecaption{Azimuthal Variations \label{azfit}}
\tablehead{
\multicolumn{1}{c}{} &\multicolumn{2}{c}{Width\tablenotemark{a}} &\multicolumn{2}{c}{Flux Density} &\multicolumn{1}{c}{Lower Limit}\\
\cline{2-5}
\colhead{} &\colhead{Amplitude} &\colhead{PA\tablenotemark{b}} &\colhead{Amplitude\tablenotemark{c}} &\colhead{$PA_{0}$} &\colhead{$g$}
}
\startdata
F110W    &2.3   &-1    &4.3 $\pm$ 0.7     &22 $\pm$ 4   & 0.39 $\pm$ 0.02 \\
F160W    &3.1   &-8    &3.1 $\pm$ 0.4     &19 $\pm$ 7   & 0.30 $\pm$ 0.02 \\
\enddata
\tablenotetext{a}{No uncertainties are calculated for the fit to the width variation 
as the fitting function is not sufficiently well matched to the data (see \S ~\ref{1-D}).}
\tablenotetext{b}{Note that the PA is for the thinnest region of function.}
\tablenotetext{c}{This is the near-to-far flux ratio (see \S\ref{azvar}).}
\end{deluxetable}

\begin{deluxetable}{lll|llll}
\tablewidth{7.0in}
\tabletypesize{\footnotesize}
\tablecaption{Comparison of Observed \& Modeled Disk Properties \tablenotemark{a}\label{nearfar}}
\tablehead{
\multicolumn{1}{l}{} &\multicolumn{2}{c}{Data} &\multicolumn{4}{c}{Models} \\
\cline{2-3}\cline{4-7}
\colhead{} &\colhead{This paper} &\colhead{R96} &\colhead{This paper} &\colhead{This paper} &\colhead{Wood\tablenotemark{b}} 
&\colhead{Wood\tablenotemark{b}} \\
\colhead{} &\colhead{} &\colhead{} &\colhead{model 1} &\colhead{model 2} &\colhead{{\it no CS disks}} &\colhead{{\it CS disks}}
}
\startdata
$R_{M_{F110W}}$               &0.014 $\pm$ 0.0003 &0.0043  $\pm$ 0.0012   &0.007  &0.012   &0.030    &0.008 \\
$\Delta(M_{F110W}-M_{F160W})$ &0.10 $\pm$ 0.03    &0.65 $\pm$ 0.32        &0.50   &0.47    &-0.07    &-0.31 \\
$\Delta(M_{F160W}-M_{F205W})$ &-0.04 $\pm$ 0.06   &-0.34 $\pm$ 0.45       &0.28   &0.20    &-0.21    &-0.20 \\
$NF_{F110W}$\tablenotemark{c} &1.41 $\pm$ 0.03    &3.6 $\pm$ 1            &1.90   &2.35    &0.67     &4.11 \\
$NF_{F160W}$              &1.39 $\pm$ 0.02        &2.6 $\pm$ 0.25         &1.82   &2.03    &0.67     &4.07 \\
$NF_{F205W}$              &\nodata                &1.5 $\pm$ 0.5          &1.70   &1.741   &0.67     &3.65 \\
\enddata
\tablenotetext{a}{Note that the R96 and Wood et al. (1999) results are based on ground-based JHK filter observations,
which differ slightly from the NICMOS filters used here (see \S\ref{obs} for details).}  
\tablenotetext{b}{Wood et al. 1999 model 1 (no CS disks) and model 3 (CS disks) results, from their Table 3.}
\tablenotetext{c}{Integrated flux from the `near' side of the disk (180$^{\circ}$ section centered on northern-most
part of the disk semi-minor axis), divided by that from the `far' side (180$^{\circ}$ section centered on southern-most
part of the disk semi-minor axis).}
\end{deluxetable}

\begin{deluxetable}{lcc|cc}
\tablewidth{2.5in}
\tablecaption{Model Dust Grain Properties\tablenotemark{a} \label{dustprops}}
\tablehead{
\multicolumn{1}{l}{$\lambda$} &\multicolumn{2}{c}{This paper} &\multicolumn{2}{c}{Wood} \\
\cline{2-5}
\colhead{(\micron\,)} &\colhead{{\bf $\omega$}} &\colhead{{\bf $g$}} &\colhead{{\bf $\omega$}} &\colhead{{\bf $g$}}
}
\startdata
1.03 &{\bf 0.47} &{\bf 0.85} &\nodata &\nodata \\
1.25 &0.44 &0.79 &0.46 &0.32 \\
1.55 &{\bf 0.40} &{\bf 0.74} &\nodata &\nodata \\
1.6  &0.39 &0.73 &0.42 &0.29 \\
1.9 &{\bf 0.35} &{\bf 0.69} &\nodata &\nodata \\
2.2 &0.31 &0.64 &0.35 &0.25 \\
\enddata
\tablenotetext{a}{Values in bold represent those used to compare the NICMOS
observations.}
\end{deluxetable}
\clearpage

\newpage
\figcaption[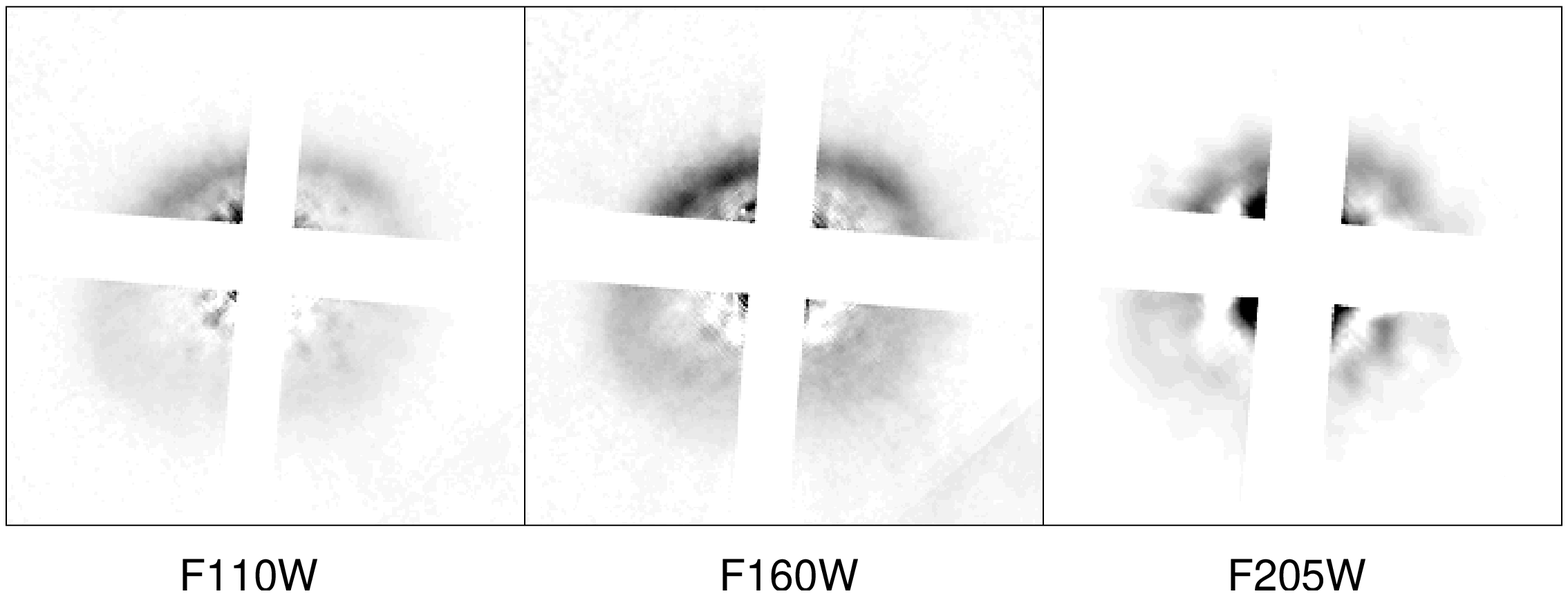]{Models of the central binary, generated through PSF fitting,
have been subtracted from the F110W, F160W and F205W images of the GG Tau system, 
revealing the large circumbinary ring. The images here show the central 
$\sim$5\arcsec\ region around the binary system GG Tau A, with North up and East
to the left. \label{mosaic}}

\figcaption[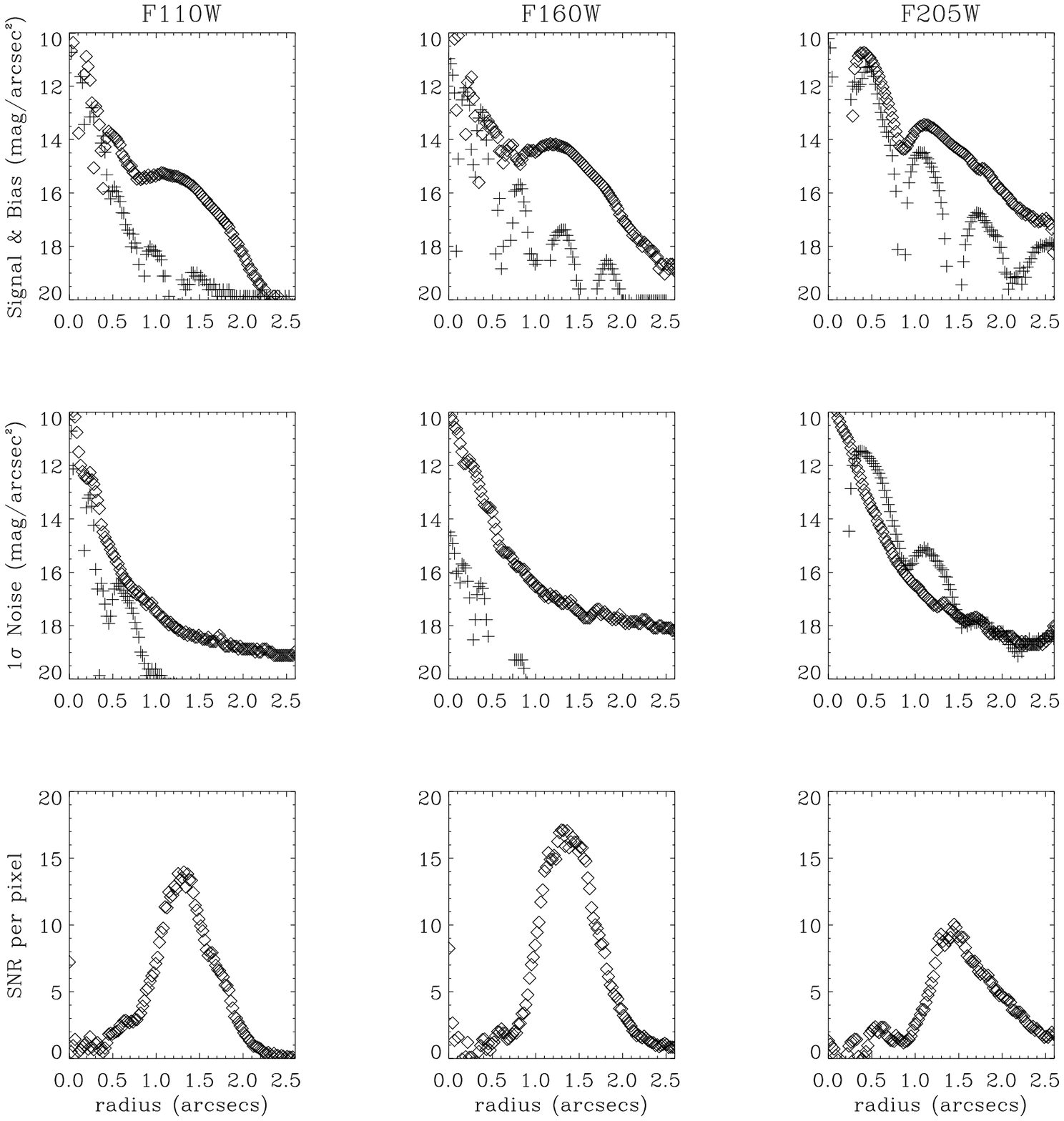]{Radial plots of detected disk emission and estimated
biases and uncertainties are shown.
The first row plots the surface brightness azimuthal averages 
of the disk light (diamonds) and bias measurements (crosses; see section \S\ref{bias}).
The second row shows the azimuthal averages for the 1$\sigma$ measurement uncertainty 
surface brightness (diamonds), and the statistical error
associated with the bias estimates (crosses). The last row shows the resulting
signal to noise for each filter, calculated using equations (1) and (2). 
The circumbinary disk is clearly detected at radii greater than $1\arcsec$, 
but the low SNR at radii less than $\sim$0\farcs8 make detection
of material in the disk gap impossible. \label{profile} }

\figcaption[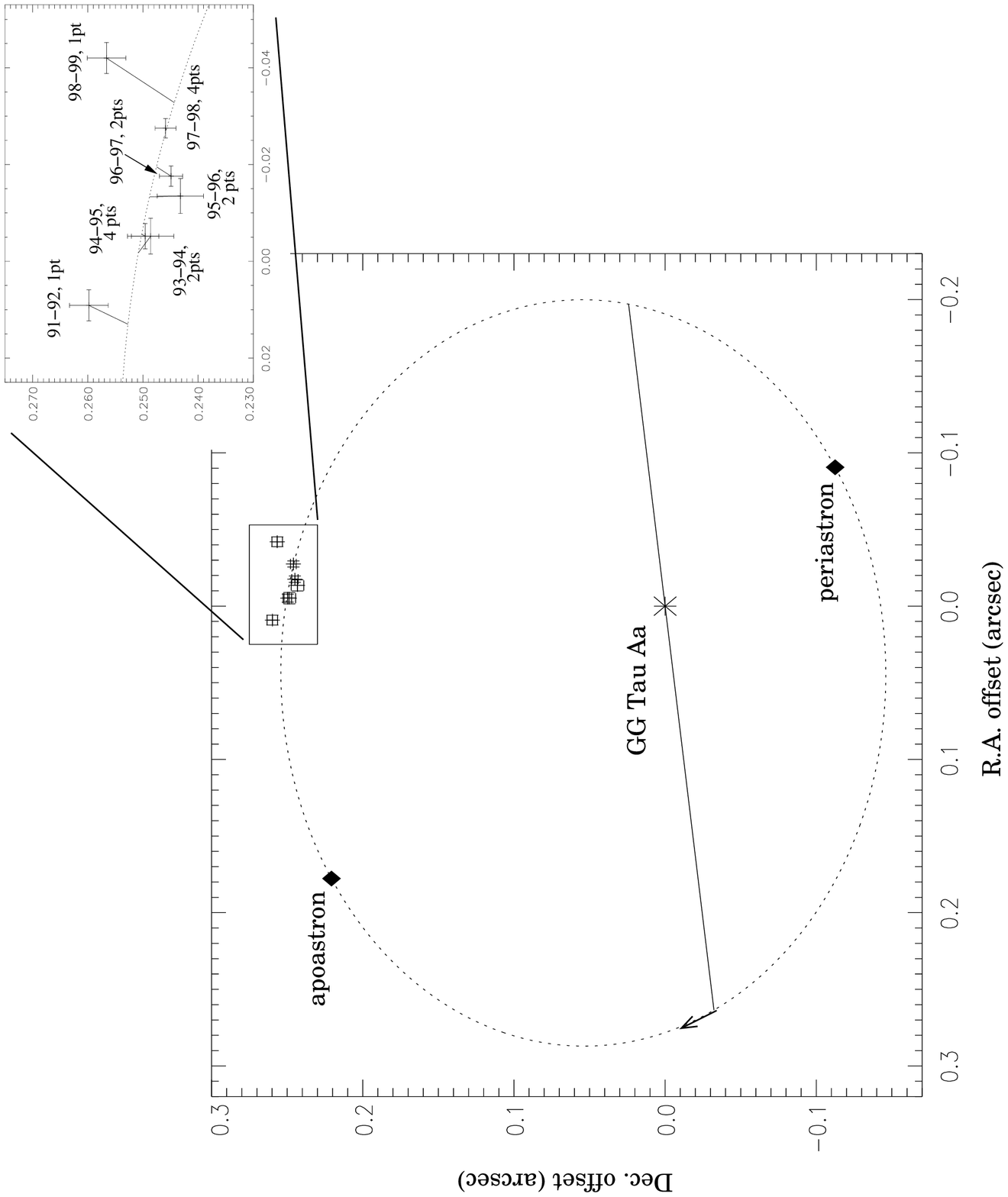]{The GG Tau orbit, projected onto the plane of the sky.
The dashed line corresponds to the line of nodes: the arrow on the right hand
node denotes the ascending node, and hence the direction of orbital motion.
Periastron and apoastron are marked.
Overplotted on the orbit is the measured offset of the secondary from GG Tau Aa
for seasonal averages of all the published data on this system, where observations 
from October through January of the next year are considered to be in the
same season. These data points are connected 
to their corresponding modeled positions by thin lines. \label{orb} }

\figcaption[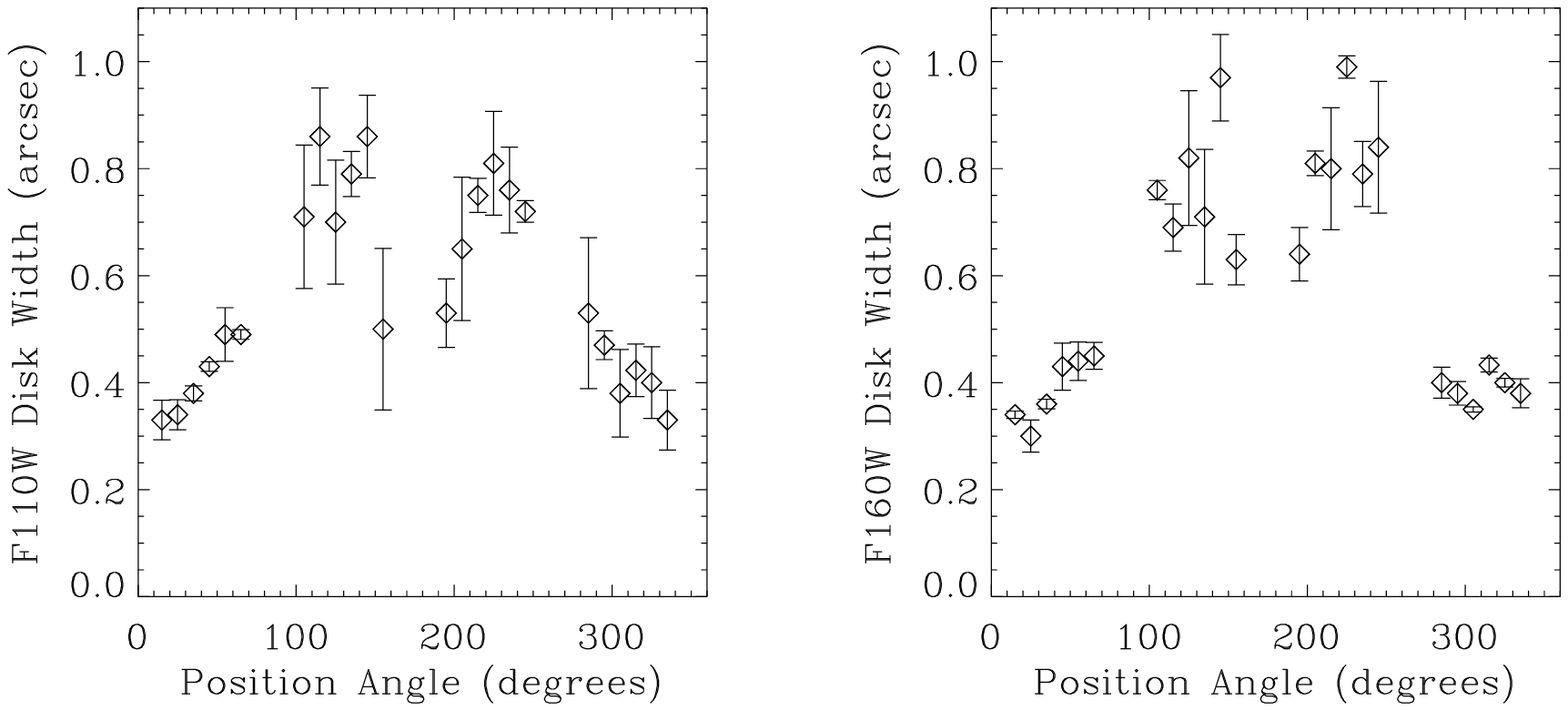]{The measured disk width, in arcseconds,  
per 10 degree sector, as a function of position angle on the sky for the 
F110W and F160W data.  \label{width} }

\figcaption[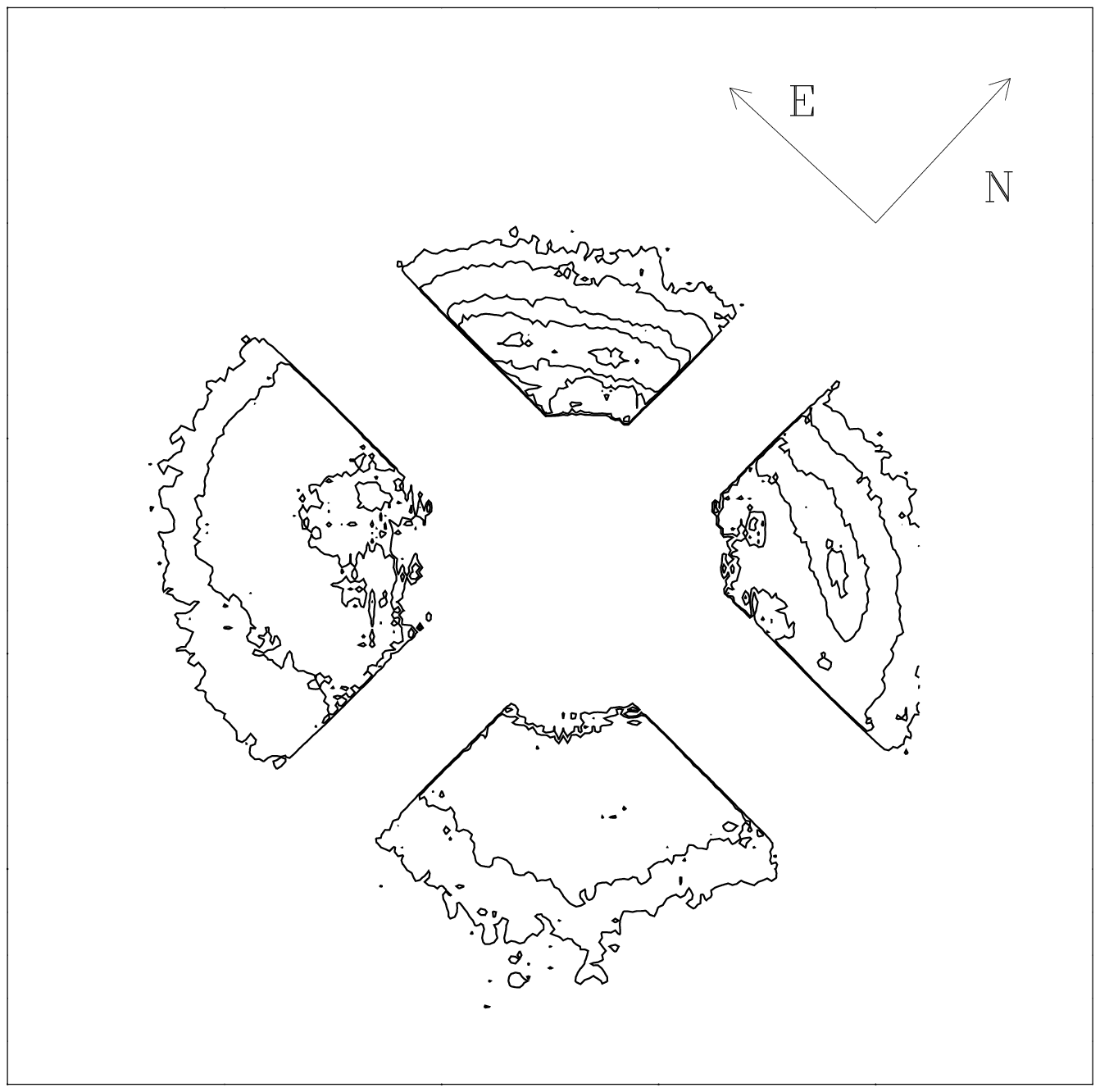]{The F160W image of the disk, shown
in contour levels set at 8\%, 16\%, 42\%, 58\% and 83\% of the
disk maximum.  The kink in the disk is evident in the south-east
quadrant of the disk where the contours turn sharply, creating an
'elbow' in the disk, and then flatten off along the southern edge of
the disk.  \label{kink} }

\figcaption[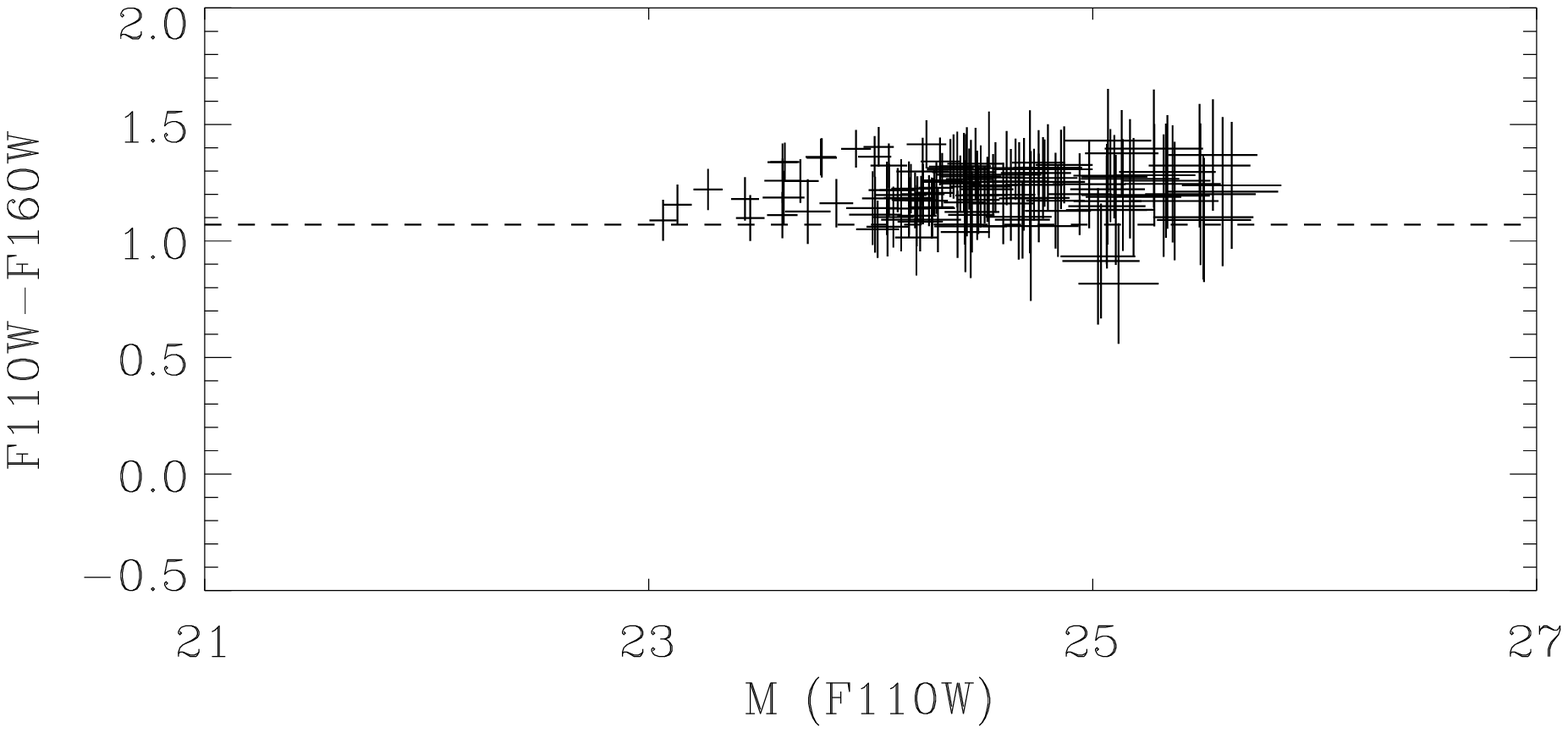]{ Variations in color across the disk
are investigated by calculating the magnitude in 9 pixel wide apertures 
placed within the disk photometric aperture. The $F110W-F160W$ 
color-magnitude diagram show
no obvious color variations around the disk. \label{diskmag} }

\figcaption[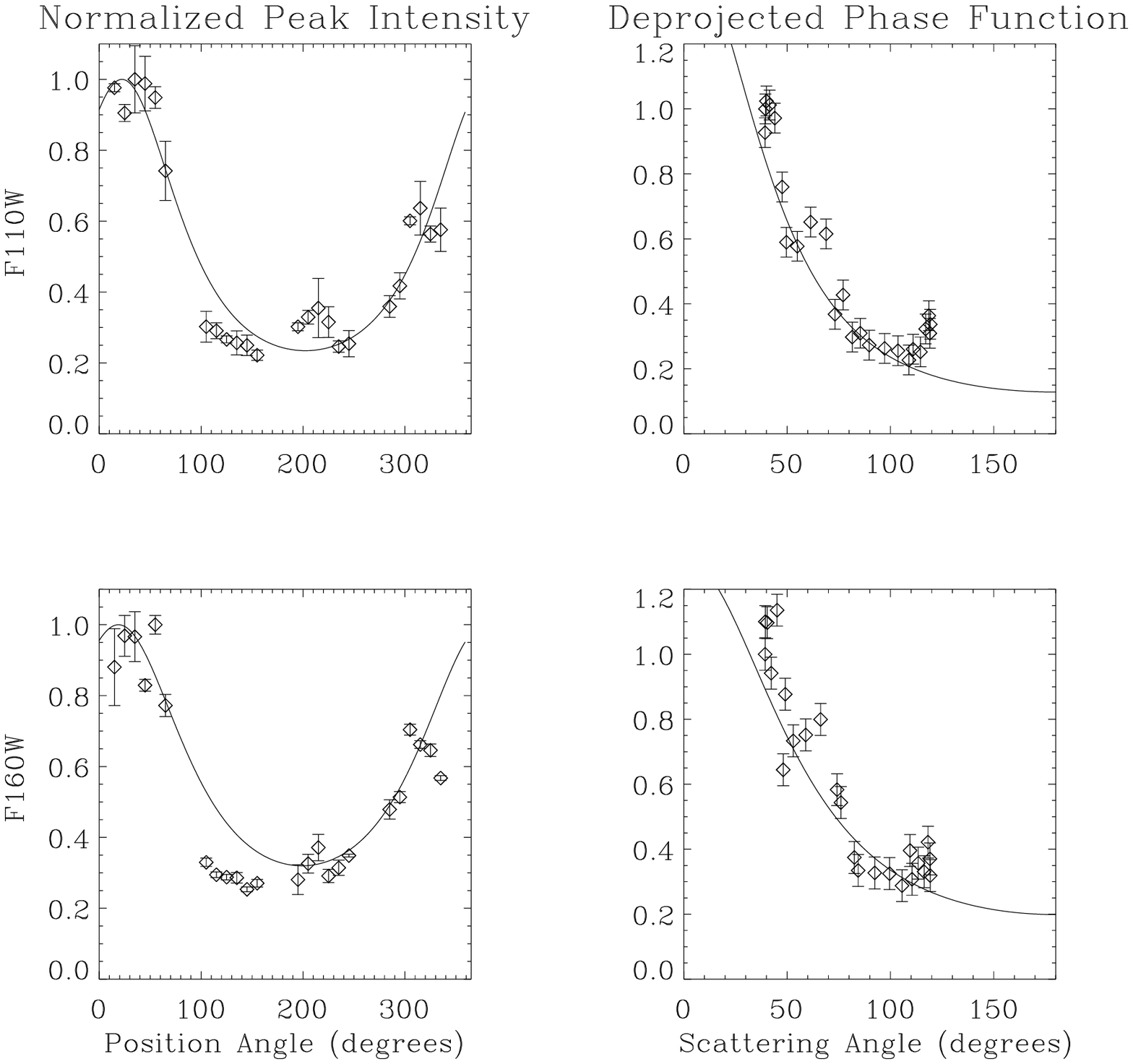]{ {\it (left)} Normalized disk peak flux 
per 10 degree sector, as a function of position angle on the sky for the 
F110W and F160W data. The Henyey-Greenstein scattering phase functions
which maximize the disk intensity variations are shown (solid lines).
The values of PA$_{0}$ found from the fitting process can be 
found in Table~\ref{azfit}. {\it (right)} The observed azimuthal flux variation
translated into a scattering phase function. The scattering phase function
is symmetric around 180$^{\circ}$ and we therefore only show the 0$\rightarrow$180
degree range. The best fitting Henyey-Greenstein scattering phase function 
for each filter is also shown (solid line), with $g = 0.39$ and $0.30$ at 1.03 \micron\  and 
1.55 \micron\ respectively. \label{phase} }

\figcaption[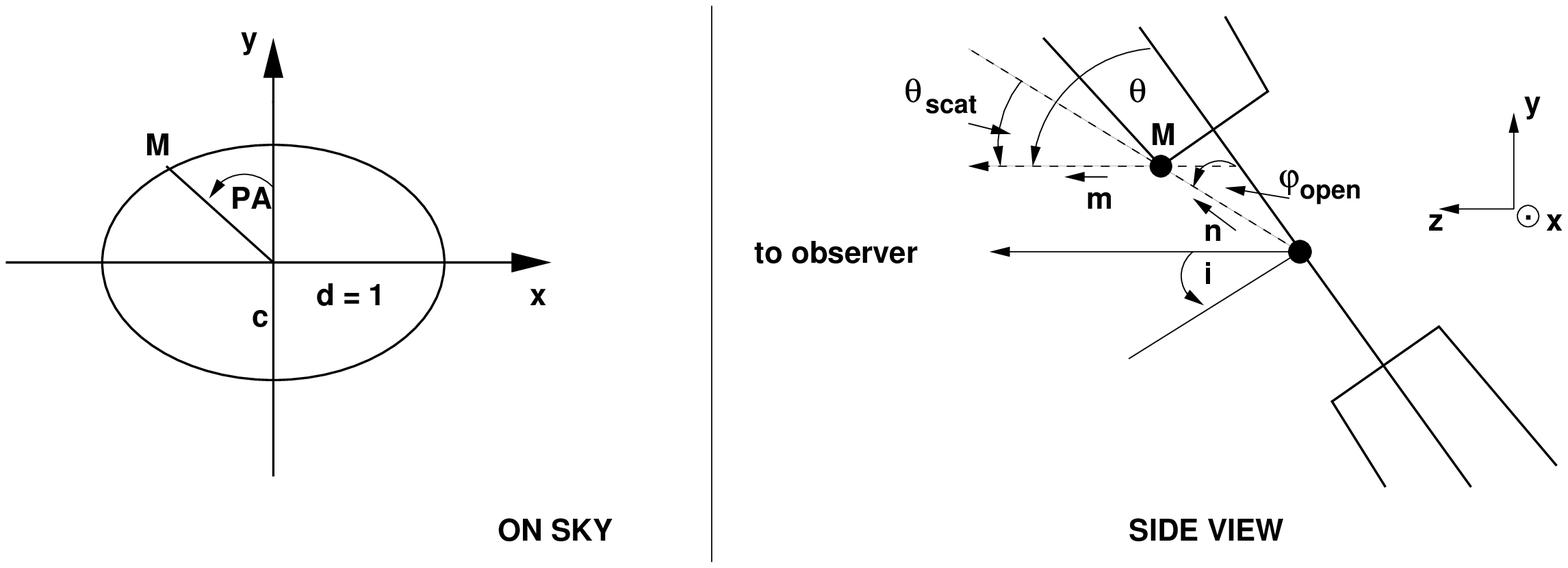]{ A cartoon of the disk showing {\it
    (left)} an on-sky view with the defined elliptical axes and
    position angle and {\it (right)} a side-view showing the relation
    between scattering angle, opening angle and disk geometry. \label{model} }

\clearpage

\begin{figure}
\figurenum{1}
\plotone{f1.eps}
\end{figure}

\begin{figure}
\figurenum{2}
\plotone{f2.eps}
\end{figure}

\begin{figure}
\figurenum{3}
\plotone{f3.eps}
\end{figure}

\begin{figure}
\figurenum{4}
\plotone{f4.eps}
\end{figure}

\begin{figure}
\figurenum{5}
\plotone{f5.eps}
\end{figure}

\begin{figure}
\figurenum{6}
\plotone{f6.eps}
\end{figure}

\begin{figure}
\figurenum{7}
\plotone{f7.eps}
\end{figure}

\begin{figure}
\figurenum{8}
\plotone{f8.eps}
\end{figure}

\end{document}